**The role of lengthscale in the creep of Sn-3Ag-0.5Cu solder microstructures**

Tianhong Gu*, Christopher M. Gourlay, and T. Ben Britton

Department of Materials, Imperial College London, SW7 2AZ. UK

*Corresponding author: t.gu15@imperial.ac.uk

**Abstract**

Creep of directionally solidified Sn-3Ag-0.5Cu wt.% (SAC305) samples with near-<110> orientation along the loading direction and different microstructural lengthscale is investigated under constant load tensile testing and at a range of temperatures. The creep performance improves by refining the microstructure, i.e. the decrease in secondary dendrite arm spacing ($\lambda_2$), eutectic intermetallic spacing ($\lambda_e$) and intermetallic compound (IMC) size, indicating as a longer creep lifetime, lower creep strain rate, change in activation energy (Q) and increase in ductility and homogeneity in macro- and micro-structural deformation of the samples. The dominating creep mechanism is obstacle-controlled dislocation creep at room temperature and transits to lattice-associated vacancy diffusion creep at elevated temperature ($\frac{T}{T_M}$ > 0.7 to 0.75). The deformation mechanisms are investigated using electron backscatter diffraction (EBSD) and strain heterogeneity is identified between $\beta$-Sn in dendrites and $\beta$-Sn in eutectic regions containing Ag$_3$Sn and Cu$_6$Sn$_5$ particles. The size of the recrystallised grains is modulated by the dendritic and eutectic spacings, however, the recrystalised grains in the eutectic regions for coarse-scaled samples (largest $\lambda_2$ and $\lambda_e$) is only localised next to IMCs without growth in size.

**Keywords:** Pb-free solder, Digital image correlation, Microstructural evolution, Polygonisation, and Recrystallisation.

# 1. Introduction

Sn-3Ag-0.5Cu wt.% (SAC305) is one of the most commonly used solder alloys in electronic interconnections [1]. SAC305 consists of primary $\beta$-Sn dendrites and eutectic regions with $Cu_6Sn_5$ and $Ag_3Sn$ embedded within the $\beta$-Sn matrix. $\beta$-Sn occupies ~ 96% of the SAC305 solder and exhibits significant anisotropy in physical properties, such as elastic stiffness and thermal expansion resulting from its body-centred tetragonal (BCT) structure [2, 3].

$Cu_6Sn_5$ and $Ag_3Sn$ exhibit greater stiffness and strength than $\beta$-Sn [4-6]. The $Ag_3Sn$ particles are considered to be the primary strengthening mechanisms in the SAC solder matrix, contributing to strength by impeding dislocation movement [7-11].

Past work has shown that the creep behaviour of a given solder alloy is affected by the initial microstructure, which can be controlled by the cooling rate [12-16] and the undercooling of $\beta$-Sn nucleation [17] during solidification; faster cooling rates and deeper undercoolings result in a smaller dendrite arm spacing, eutectic IMC size and spacing. An increase in creep lifetime, secondary creep strain rate and ductility have been reported for samples cooled at a faster rate [12-17]. Among these reports, Wu et al. [12] found that the creep life and creep strain rate of their tested Sn-3.5Ag dog-bone samples changed about 1.5 and 1-2 orders of magnitude respectively between fast and slow cooled samples. Less creep resistance and a reduction in ductility for Sn-Ag/Cu joints solidified at a slower rate were indicated by Yang et al.[13], who reported it is caused by the brittle nature of the IMCs. Kim et al. [14] pointed out that, under tensile testing SAC305, Sn-3.5Ag-0.7Cu and Sn-3.9Ag-0.6Cu samples formed using faster cooling rate have a greater elongation and ductility with decreasing size of $Ag_3Sn$. Moreover, Ochoa et al. [15] observed a slight decrease of the activation energy, $Q$ from 40 to 35 kJ/mol by increasing the cooling rate from 0.08 to 0.5 °C/s for Sn-3.5Ag bulk solder.

Aging of the solder microstructure can also impact the creep behaviour of solder alloys [18-24] and similar to the solidification studies, the creep strain rate increases with larger IMC size [18-24]. Dutta et al. [18] studied Sn-3.5Ag and Sn-4Ag-0.4Cu and concluded that creep is controlled by dislocation climb and the rate is dependent on the IMC size, with evidence of climb at the IMC/Sn interface. In comparing aged and rolled and cut 95.5Sn-3.9Ag-0.6Cu samples, Vianco et al. [20, 21] found a decrease in both stress exponent ($n$) and activation energy ($Q$) from aged to rolled and cut samples for temperatures in the range of -25 – 160 °C. This was different to Basit et al. [24] and Talebanpour et al. [25] who found no change in $Q$ as the aging time and temperature increased for pure Sn, SAC105 and SAC305 solders tested between -25 and 125 °C.

Plastic deformation is heterogenous for Sn-based solders, resulting from an increase in stored energy and dislocation-dislocation interactions. The stored energy in the crystal lattice is reduced by deformation phenomena, namely polygonisation and recrystallisation, and these are precursors to crack nucleation and propagation in the strain-concentrated regions [26-31].

In service, the SAC305 solder works at a sufficiently high homologous temperature (> 0.6) that deformation results in microstructural evolution, including dynamic recovery and recrystallisation. During recovery processes the solders deform by continuous lattice and rigid body rotation, where the stored energy or the driving force can be released by forming subgrains with misorientation between 2 - 15° [27-29, 32]. With increasing strain, the microstructure of the solder develops continuously and is often found to recrystallise by forming high angle grain boundaries (HAGBs). Discontinuous recrystallisation can also take place because of the presence of grain boundaries and second phase particles (IMCs), which act as obstacles to dislocation movement and cause localised dislocation pile-up [30, 31, 33-36].

While there is important ongoing work on the macroscopic creep response of solders [37-40], there is also a need for studies that account for crystallographic orientations and the microstructure at both the $\beta$-Sn dendrite scale and the $Ag_3Sn$ and $Cu_6Sn_5$ particle scale. The studies above did not include information on crystallographic orientation and microstructural evolution during creep deformation, which may play a significant role. It has been shown by Ma and Suhling [41] that there are large discrepancies in the measurement of mechanical properties for solders between studies and these differences result from the microstructural variations between the tested samples.

In the present work, we simplify the microstructure using directional solidification (DS) to reproduce samples with the same microstructure and controlled crystal orientation and select the microstructural lengthscale through controlling the growth rate. In this work, we explore the effect of secondary dendrite arm spacing ($\lambda_2$), eutectic IMC spacing ($\lambda_e$) and IMC size on the creep behaviour and microstructural evolution for constant stress creep testing at a range of temperatures. Changes on creep behaviour are explored mechanistically with studies of strain accumulation and recrystallisation (location and magnitude) at different microstructural lengthscales by using two-dimensional digital image correlation (2D DIC) and electron backscatter electron diffraction (EBSD).

## 2. Sample microstructure and experimental procedure

Dog-bone samples (*Figure 1*a) were produced by DS, using the method given in ref [42-44], with a 350 °C hot zone and 25 °C cold zone and three different pulling rates: 2, 20 and 200 µm/s. A constant pulling rate through a near-constant positive temperature gradient produces a uniform $\lambda_2$, $\lambda_e$, and eutectic IMC ($Ag_3Sn$ and $Cu_6Sn_5$) size, and their lengthscale is controlled by the pulling velocity. This also results in large <110> oriented $\beta$-Sn dendrites along the long axis during growth. Together these are required for the mechanistic understanding presented in this work.

The three growth velocities of 2, 20 and 200 µm/s produced three different microstructural lengthscales as shown in *Figure 1b - 1j* and measurements of these are collated in *Table I*. The spacing metrics, $\lambda_2$ and $\lambda_e$, decrease by approximately an order of magnitude as the growth velocity increases from 2 to 200 µm/s. The eutectic $Ag_3Sn$ (lighter features) and $Cu_6Sn$ (darker features) can be distinguished according to their atomic weight (*Z*) in the backscattered electron (BSE) images in *Figure 1e - 1g*. Furthermore, *Figure 1e - 1j* show that the eutectic $Ag_3Sn$ undergoes a plate-to-rod transition with increasing growth rate between 2 and 20 µm/s, consistent with [45-47]. The eutectic $Cu_6Sn_5$ has a rod-like morphology for all three microstructures and a decreasing lengthscale with increasing growth velocity *(Figure 1h - 1j)*. Note that the medium microstructure sample was published in ref [42].

| Initial Microstructure | v (µm/s) | $\lambda_2$ (µm) | $\lambda_e$ (µm) | $\bar{r}_{IMC}$ (µm) |
|---|---|---|---|---|
| Fine | 200 | 20 ± 3 | 0.68 ± 0.02 | 0.20 ± 0.01 |
| Medium | 20 | 50 ± 3 | 1.18 ± 0.02 | 0.63 ± 0.01 |
| Coarse | 2 | 120 ± 4 | 7.95 ± 0.05 | 2.15 ± 0.01 |

*Table I. Comparison of secondary dendrite arm spacing ($\lambda_2$), eutectic spacing ($\lambda_e$) and IMC size (radius $\bar{r}_{IMC}$) for the three DS samples. The error is the standard deviation (STD) of the mean from four similar micrographs.*

The tensile sample surface was polished before testing using broad ion beam (Gatan PECS II) after mechanical polishing (0.05 µm colloidal silica) to improve the quality of surface finish for EBSD and to generate long strip speckle patterns for DIC [42].

The samples were deformed at a constant stress of 30 (± 2) MPa at temperatures between 298 and 473 (± 6) K using a 2 kN Gatan Mtest 2000E stage. The tests were interrupted shortly after the onset of secondary creep (~120 s) and stopped at the end of tertiary stage creep.

EBSD scans were performed using Bruker e-Flash$^{HR}$ detector in a FEI Quanta SEM. The sample was then replaced within the loading frame to resume the creep test between the two EBSD scans without polishing. The EBSD maps were scanned with a step size of 6 µm and 0.4 µm [42].

Creep strain was measured by 2D optical DIC using DaVis (LaVision), from which the surface strain field was extracted and the average value of each strain field map was recorded for each time-step. A subset size of 16 × 16 pixels, window size of 65 pixels and field of view of 10 × 2 mm were set to achieve a strain map with an effective pixel size of 11.5 µm/pixel [42].

## 3. Creep behaviour of DS SAC305 solder

Maps of crystal orientation, depicted using inverse pole figure colouring with respect to the loading direction (IPF-LD), show that the initial crystal orientations in both the dendrite and eutectic *β*-Sn regions, i.e. a dark blue colour in the representative orientation maps, have a deviation angle of 3°, 12° and 7° from [110] with respect to the loading direction for the sample with coarse (*Figure 3a*), medium (*Figure 3f*) and fine (*Figure 3k*) lengthscales respectively.

Mechanical testing data is reported in *Figure 2*. The creep curves in *Figure 2*a reveal changes in creep response as a function of lengthscale. Samples with a finer IMC/dendrite lengthscale have: lower strain at the onset of secondary creep (*Figure 2*b); a lower secondary creep rate (*Figure 2*c); and longer creep lifetime (*Figure 2*d).

The creep curve data is analysed using a simple constitutive model:

$$\dot{\varepsilon} = A\sigma^n \cdot \exp\left(-\frac{Q}{RT}\right) \tag{1}$$

where R is the ideal gas constant of 8.314 J·Kmol$^{-1}$, A is a constant, n and *Q* are measured through the slope of ln ($\dot{\varepsilon}$) vs. ln (σ) and ln ($\dot{\varepsilon}$) vs. 1/T plot respectively by rearranging equation (1).

Analysis of the creep data across different temperatures using linear fitting is presented in *Table II*. Exploring the high and low temperature creep data, shown in *Figure 2*e, indicates a 'nose' that separates these two temperature domains highlighting a change in creep mechanism, consistent with prior work [20, 21, 42, 48, 49]. At low temperatures a climb-controlled dislocation mechanism operates and at higher temperatures potentially this is controlled by lattice-associated vacancy diffusion [42, 48, 49]. *Table II* collates that the activation energy, *Q*, decreases for samples with a finer lengthscale in the low temperature range and slightly increase in the high temperature range, and this is supported by *Table III* which indicates the lengthscale for each of these mechanisms.

|  | Temperature range, T (K) | Initial Lengthscale | | |
|---|---|---|---|---|
|  |  | **Fine** | **Medium** | **Coarse** |
| **Q (kJ/mol)** | $T_{av}$ | 44 | 47 | 17 |
|  | $T_{low}$ | 34 | 41 | - |
|  | $T_{high}$ | 9 | 6 | - |
| **$R^2$** | $T_{av}$ | 0.79 | 0.81 | 0.99 |
|  | $T_{low}$ | 0.99 | 0.95 | 0.97 |
|  | $T_{high}$ | 0.99 | 0.99 | 0.87 |

*Table II. Comparison of activation energy, Q between the three samples with different lengthscales at low ($T_{low}$) and high ($T_{high}$) temperature ranges and average of the tested temperature range ($T_{av}$). $T_{av}$ = 298 - 473 K, $T_{low}$ = 298 - 333 K, and $T_{high}$ = 363 - 473 K. $R^2$ is the square of the correlation coefficient of the corresponding liner fit for the ln (secondary creep strain rate) vs. reciprocal temperature curves in Figure 2e.*

| T [K] | Strain at onset of secondary creep (%) | | | Secondary creep strain rate (s$^{-1}$) | | |
|---|---|---|---|---|---|---|
|  | **Fine** | **Medium** | **Coarse** | **Fine** | **Medium** | **Coarse** |
| 298 | 0.85 | 1.78 | 2.42 | 7.71×10$^{-7}$ | 3.40×10$^{-6}$ | 1.06×10$^{-5}$ |
| 318 | 0.50 | 0.83 | - | 1.71×10$^{-6}$ | 7.21×10$^{-6}$ | - |
| 333 | 0.28 | 0.47 | 0.94 | 2.96×10$^{-6}$ | 1.35×10$^{-5}$ | 2.60×10$^{-5}$ |
| 363 | 0.16 | 0.33 | - | 8.98×10$^{-6}$ | 2.05×10$^{-5}$ | - |
| 393 | 0.10 | 0.29 | 0.40 | 1.13×10$^{-5}$ | 2.58×10$^{-5}$ | 5.86×10$^{-5}$ |
| 423 | 0.03 | 0.17 | - | 1.35×10$^{-5}$ | 2.67×10$^{-5}$ | - |
| 453 | 0.02 | 0.11 | - | 1.54×10$^{-5}$ | 2.91×10$^{-5}$ | - |
| 473 | 0.01 | 0.01 | - | 1.74×10$^{-5}$ | 3.44×10$^{-5}$ | - |

*Table III. Summary of SAC305 mechanical creep samples with three different lengthscales under fixed load of 30 MPa testing from 298 to 473K showing the values of secondary creep strain rate, strain at the onset of secondary creep. (The high temperature creep data of each lengthscales is included in supplementary Figure S2-3 and in ref [42]).*

The change in slope for the ln(secondary creep strain rate) vs. reciprocal temperature curves happens at a homologous temperature of $\sim \frac{T}{T_M} = 0.74 \pm 0.02$ for the fine-scaled samples ($\lambda_2 \approx 20 \pm 3$ μm and $\lambda_e \approx 0.68 \pm 0.02$ μm). This shifts to a lower temperature of ~ $\frac{T}{T_M} = 0.70 \pm 0.02$ for the medium-scaled samples ($\lambda_2 \approx 50 \pm 3$ μm and $\lambda_e \approx 1.18 \pm 0.02$ μm )[42]. For the coarse-scaled samples ($\lambda_2 \approx 120 \pm 4$ μm and $\lambda_e \approx 7.95 \pm 0.05$ μm), no transition is observed from the three data points measured within the testing temperature range, so it is insufficient to determine whether there is a transition in creep mechanism or not. While the intersection of two linear fits has been used to assess this mechanism change, it is more likely that the transition between mechanisms has variation with a range of $\frac{T}{T_M} = 0.70$ to $0.75$ for fine and medium-structured samples, which is estimated through error analysis.

## 4. Macroscopic evolution of strain field and microstructure

The DIC strain field figures at the onset of secondary creep (*Figure 3b, 3g, 3i*) show that the samples deformed relatively homogenously from primary to secondary creep. Strain localisation is observed within the samples and shown as hot spots in *Figure 3b, 3g, 3i* at the early stage of secondary creep.

The spatial distribution of strain heterogeneity revealed in the DIC maps is not correlated with the microstructural unit sizes. However, differences as a function of lengthscale in strain prior to secondary creep can be observed i.e. $\varepsilon_{coarse} = 2.4\% > \varepsilon_{medium} = 1.8\% > \varepsilon_{fine} = 0.85\%$ (*Figure 3b, 3g, 3i)*.

As the strain continues (*Figure 3c, 3h, 3m*), instability develops within the gauge section and flow localisation develops (highlighted in the red square in *Figure 3c, 3h, 3m*). This continues to develop in secondary and beyond into tertiary stage creep where necking forms (*Figure 3d, 3i, 3n*) and this is where fracture occurs.

The fine-scaled sample has greater reduction in cross-sectional area within the necked region combined with the greatest total elongation. This indicates that the finest lengthscale stabilises hardening and promote an increase in ductility.

EBSD-based orientation mapping (*Figure 3e, 3j, 3o*) reveals that crystal lattice orientation spreads and rotates during the deformation, and larger changes are found near the fracture surface. Recrystallisation and new grains are found in the highly strained regions in the neck (*Figure 3e)* and near the fracture surface (*Figure 3j, 3o*). The size of the formed recrystallised grains decreases significantly when the microstructure is refined (*Figure 3e, 3j, 3o*).

The variation in strain heterogeneity is illustrated in *Figure 4a, 4c, 4e* (extracted from *Figure 3*c - *d, 3h - i, 3m -n*). The strain level in the highly-strained regions (necked regions in *Figure 3d, 3i, 3n*) is significantly higher than the average strain level across the samples, while the uniformly-deformed regions have much lower strain level, and these have not reached tertiary creep at the end of the tests.

The corresponding pole figures (PFs) in *Figure 4b, 4d, 4f* show that the main orientations of the samples change gradually by rotation likely with a [001] rotational axis during creep. The (100)[010] is most likely to be active and this is consistent with lattice rotation about [001] for constrained deformation in a tensile test. This is inferred from consideration of the macroscopic loading direction, the crystal orientation, and the critical resolved shear stress ratios estimated from Zamiri et al. [50] (Calculation of Schmid factors are presented in supplementary *Table SIV*).

## 5. Microscopic evolution of microstructure

*Figure 5* shows the micrographs within the uniformly-deformed regions of each sample in the as-solidified condition. The FSD images indicate the contrast between dendrite and IMC-containing eutectic regions (*Figure 5*a, 5h, 5o). The IMCs, $Ag_3Sn$ and $Cu_6Sn_5$, are the protruding features in the eutectic regions. The $\lambda_2$, $\lambda_e$ and size of the IMCs decrease significantly from coarse- to fine-scaled microstructures (as noted in *Table I*).

EBSD mapping indicates the evolution of the lattice orientation and the lattice misorientations with increasing strain within the (macroscopically) uniformly straining regions (*Figure 5*) and the necked region (*Figure 6*, *Figure 7*).

In all microstructures, the EBSD data reveals that heterogeneity of lattice misorientations develops depending on the presence of the IMCs, and the range of this heterogeneity (by comparing the misorientation to average maps in *Figure 5e, 5l, 5s* to *Figure 5f, 5m, 5t*) is controlled by the size and distribution of the IMCs. This is reasonable as the IMCs are elastic and hard as compared to the matrix. Subgrains are observed in $\beta$-Sn around the IMCs for the coarse-scaled microstructure, near the dendrite-eutectic interfaces for the medium- and fine-scaled microstructures and at grain boundaries for the fine-scaled microstructure (indicated with red arrows in *Figure 5f, 5m,5t*). There is no obvious orientation change in the IPF-LD maps for the coarse- and medium-scaled microstructures from the as-solidified condition (*Figure 5b, 5i)* to the onset of secondary creep (*Figure 5c, 5j*). For the fine-scaled microstructure, the number of pink grains increases and they grow in size within the IPF-LD map (indicated with red arrows in *Figure 5q*).

At the end of tertiary stage creep (*Figure 5d, 5g, 5k, 5n, 5r, 5u*), within the mapped regions the lattice rotation [32, 51] is more obvious in the sample with a finer microstructure (*Figure 5g, 5n, 5u*) because the evenly distributed fine-scaled obstacles (e.g. IMCs and dendrite-eutectic boundaries) result in increase in total stored energy of the sample and shown as polygonisation in *Figure 5d, 5k, 5r*). In the coarse-scaled microstructure the average orientation does not change (*Figure 5d*) but significant heterogeneity is observed near the IMCs (*Figure 5g*).

In all samples the subgrain structures start to grow into the primary $\beta$-Sn dendrites, and the size of the subgrains is controlled by the size and spacing of IMCs (*Figure 5g, 5n, 5u*), further explanation is referred to Discussion 6. Furthermore, the magnitude of the misorientations in the fine-scaled subgrains is larger, which correlates with the total strain developed in each sample (*Figure 4a, 4c, 4e*). The magnitude of the misorientation here is related to the stored energy and this hints at why recrystallisation, and ultimately failure, is changed by the IMC and dendrite lengthscale.

*Figure 6* shows the micrographs of the samples within the necked region at the end of tertiary creep. The FSD images in the mapped regions (*Figure 6a - 6c*) show that the dendrite arms are elongated along the loading direction, consistent with the major strain axis. In these regions, local recrystallisation (indicated with white arrows in *Figure 6d - 6f*) is observed together with regions of high misorientation (indicated with black arrows in *Figure 6g - 6i*) and these domains are controlled by $\lambda_2$ and $\lambda_e$.

The lengthscale connection between these and recrystallisation is confirmed in the necked region, where additionally increased surface roughness (*Figure 7a - 7c*) and the associated 'rainbow' recrystallised grains (*Figure 7d - 7f*) are observed at the fracture surface. The small recrystallised grains are located only around the IMCs for the coarse-scaled microstructure, which are highlighted with blue circles in *Figure 7d*. Three large recrystalised grains are formed in the *β*-Sn dendrite for the coarse-scaled microstructure (*Figure 7d*). For the medium-scaled microstructures, the new recrystalised grains in the dendrites (green) are rotated towards [100] orientation, which is the loading direction (indicated with white arrows in *Figure 7e*). In the eutectic region, the recrystalised grains (yellow) are rotated towards [001] orientation (indicated with black arrows in *Figure 7e*). For the fine-scaled microstructure, the 'rainbow' recrystalised grains are formed in dendritic *β*-Sn (*Figure 7f*), which deforms by gradual lattice rotation with continuous development of polygonisation and causes recrystallisation in the highly strained region [42]. The constrained stored energy is released at the fracture surface showing decrease in misorientation (*Figure 7g - 7i)*.

## 6. Discussion: The role of lengthscale on creep mechanisms

In secondary creep, the two competing processes of strain hardening and dynamic recovery are in balance and no localised creep damage is obtained in the macroscopic scale (*Figure 2*a). As the sample starts to yield, the high strain gradient regions are generated by the IMCs and/or dendrite - eutectic interfaces and create unstable regions through the gauge section (DIC strain field maps in *Figure 3b, 3g, 3i)* and change the accumulation of stored energy (shown with the EBSD maps in *Figure 5*c, 5f, 5j, 5m, 5q, 5t). These observations highlight how we can optimise the microstructure, through a change in microstructural lengthscale to adjust creep life and creep rate.

During tertiary creep (*Figure 2*a), there is a substantial increase in creep strain with time, resulting in necking and fracture. The necked region is important as this is where failure occurs, which has motivated the study here of the microstructural contributions to accumulated strain and failure. If the neck is stabilised, through creep strain hardening, not only the ductility is high but also results in slower secondary creep and a longer creep life. Ultimately, failure of the sample does occur at the neck resulting in the localisation of creep strain and surface roughening appear with formation of stable necks (*Figure 3d, 3i, 3n*).

Failure of the fine-scaled sample results in a sharper neck (*Figure 3o*) as the total amount of strain in the neck is large before the onset of tertiary creep (*Figure 2a*). The transition to tertiary creep is more spatially spread for the fine-scaled sample with a much larger necked region (*Figure 3o*) than the coarse-scaled sample (*Figure 3e*). This supports the idea that there is less instability, and the volume of material that recrystallised is smaller. The local recrystallisation is more prevalent in the fine-scaled sample, where the fine-scaled and evenly-distributed obstacles (IMCs and dendrite-eutectic boundaries) can restrict dislocation motion and grain boundary mobility easily and cause interaction between two slips to form substructure (subgrains and recrystalised grains), as evidenced with lattice gradient in *Figure 5 - Figure 7*. On the other hand, the slip happens just around the IMCs for the coarse-scaled sample due to the large size of and heterogeneously-distributed IMCs, causing the recrystallisation within more localised neck and failure of the sample happens sooner in time. Thus, the ductility of the SAC305 sample increases with refining the microstructural lengthscale.

In the present work, the ex-situ tests with repeat imaging of the same area indicate that recrystallisation occurs during deformation and this creates crystallographic texture. This is supported by our recent prior work [42]. The present work is critical as it highlights that the establishment of strain gradients is near the IMCs during deformation (*Figure 5*). The strain gradients result in local subgrains which can be thought of as regions of low dislocation content separated by dislocation walls, resulting in substantive changes in lattice orientation. These subgrains store energy as the deformation progresses and likely act as nucleation sites for recrystallisation. This is associated with the concept of particle stimulated nucleation (PSN) [52].

*Figure 8g, 8h* quantify the change in size of recrystallised grains within dendrite and eutectic regions separately for the three samples with different microstructural lengthscales (*Figure 5 - Figure 7*). Due to the constraint of $\lambda_2$, $\lambda_e$ and IMC size, the recrystallised grains in the eutectic *β*-Sn regions have much smaller grain size than in the *β*-Sn dendrites for all three microstructural lengthscales (*Figure 8*), as also described in ref [42].

In addition to recrystallisation around the IMCs, recrystallisation can occur within the dendrite. This is important for the coarse-scaled sample, where the IMCs are fewer and more widely spaced, so the recrystalised grains propagate relatively easily and are quick to deform through the *β*-Sn matrix in the dendrite (annotated in *Figure 8d*). In this sample, the creep failure is related to propagation of the recrystallisation bands in the *β*-Sn dendrites (*Figure 3*e). This is related to where the neck and the ultimate crack form.

As the microstructural lengthscale becomes coarser, there is a significant strain partitioning between the primary *β*-Sn dendrites and the eutectic regions (*Figure 5 - Figure 7*). The macroscopic deformation of the sample (*Figure 3*) becomes less stable once the recrystallisation bands start to form and these recrystallisation bands extend relatively quickly through the entire gauge section being evidenced by significant strain localised within the necked region, and in turn during the final stages of necking and failure, this leads to a large volume of recrystallisation (*Figure 3*e, 3j, 3o).

In the present work, our macroscopic observations of strain (see *Figure 3*) and our microscopic observations of lattice rotations and recrystallisation (see *Figure 5 - Figure 7*) indicate that the IMC size within the eutectic controls the secondary stage creep to the onset of tertiary stage creep. As it is challenging to isolate IMC size and the distribution of eutectic and primary Sn, we must note that we cannot rule out a further effect due to the 'patch size' of these different domains and this would be an interesting topic to consider in further work, perhaps via a coarsening study or with related crystal plasticity simulations.

**7. Conclusions**

Creep strain patterning, stored energy accumulation, recrystallisation and ultimately failure of SAC305 solders are controlled by the size and distribution of IMCs and the size of dendrites. This has been studied using reproducible samples with controlled microstructural lengthscales, namely crystal orientation (close to [110] and [100]), secondary dendrite arm spacing ($\lambda_2$), eutectic spacing ($\lambda_e$) and $Ag_3Sn$ and $Cu_6Sn_5$ size. The following conclusions can be drawn from this work.

1. The creep mechanisms change with microstructural lengthscale of the microstructure within sample. Longer creep lifetime, lower secondary creep strain rate and lower strain at onset of secondary creep are obtained by refining the lengthscale. The activation energy, Q decreases in the low temperature range (298 - 333 K) and increases in the high temperature range (363 - 473 K) with refining lengthscale. Deformation changes the mechanism at higher temperature and this is likely a transition from climb-controlled dislocation to lattice-associated vacancy diffusion creep. These behaviours imply that creep deformation is obstacle-controlled, which becomes more prominent below the transition temperature.

2. As the microstructural lengthscale changes, the microstructure of the samples evolves differently at the macroscopic scale. Heterogeneous deformation occurs during creep. The sample with a finer lengthscale forms a more stable neck than the sample with a coarser lengthscale. The ductility of the sample increases with refining lengthscale and the deformation becomes less localised, i.e. more homogenous deformation is introduced.

3. At the microscopic scale, the heterogeneous evolution of microstructure is caused by the presence of two distinct microstructural regions, i.e. primary $\beta$-Sn dendrites and IMC-containing eutectic $β$-Sn regions as described in ref [42]. The initial deformation starts in the $β$-Sn within the eutectic region near IMCs because dislocations often concentrate against the hard particles (IMCs), which becomes highly localised around the IMCs for a coarse-scaled sample, whereas more spatially extensive deformation for a fine-scaled sample. The soft phase, $β$-Sn, deforms by lattice rotation to form subgrains with continuous development of misorientation (polygonisation) and generates recrystallisation with large accumulation of strain at tertiary creep and this is enhanced with a finer microstructural lengthscale. The size of the formed subgrains and recrystalised grains decreases with increasing lengthscale of the sample, i.e. polygonisation and recrystallisation are controlled by $λ_2$ and $λ_e$.

**Author Contributions:**

TG drafted the initial manuscript and conducted the experimental work. CG and TBB supervised the work equally. All authors contributed to the final manuscript.

**Acknowledgements:**

TBB would like to thank the Royal Academy of Engineering for his research fellowship. We would like to thank EPSRC (EP/R018863/1) for funding. We would like to acknowledge Dr Sergey Belyakov for support in the initial fabrication of the samples. The assistance of Dr Te-Cheng Su in DIC studies is also gratefully acknowledged. The microscope and loading frame used to conduct these experiments was supported through funding from Shell Global Solutions and is provided as part of the Harvey Flower EM suite at Imperial.

**Conflict of Interest:**

The authors declare that they have no conflict of interest.

**References**

1. R.J. Coyle, Sweatman, K. and Arfaei, B., *JOM,* 67, 2394; (2015).
2. A.U. Telang and T.R. Bieler, *Research Summary Lead-Free solder,* 44; (2005).
3. J.W. Xian, G. Zeng, S.A. Belyakov, Q. Gu, K. Nogita and C.M. Gourlay, *Intermetallics,* 91, 50; (2017).
4. E. S. Hedges, *Tin and its alloys: by Ernest S. Hedges*: Arnold, (1960).
5. V.M.F. Marques, C. Johnston, and P.S. Grant, *Acta Mater.,* 61, 2460; (2013).
6. D. Li, C. Liu and P.P. Conway, *Mater. Sci. & Engineer.: A,* 391, 95; (2005).
7. S. Choi, J.G. Lee, F. Guo, T.R. Bieler, K.N. Subramanian and J.P. Lucas, *JOM,* 53, 22; (2001).
8. F.X. Che and J.H.L. Pang, *J. Alloys Compd.,* 541, 6; (2012).
9. R. Parker, R. Coyle, G. Henshall, J. Smetana and E. Benedetto, *J. Proc. SMTAI,* 348; (2012).
10. R. Coyle, P. Read, H. McCormick, R. Popowich and D. Fleming, *J. SMT,* 25, 28; (2011).
11. S. Terashima, Y. Kariya, T. Hosoi and M.Tanaka, *J. Electron. Mater.,* 32, 1527; (2003).
12. K. Wu, N. Wade, J. Cui and K. Miyahara, *J. Electron. Mater.,* 32, 5; (2003).
13. W. Yang, L.E. Felton and R.W. Messler, *J. Electron. Mater.,* 24, 1465; (1995).
14. K.S. Kim, S.H. Huh and K. Suganuma, *Mater. Sci. & Engineer.: A,* 333, 106; (2002).
15. F. Ochoa, X. Deng and N. Chawla, *J. Electron. Mater.,* 33, 1596; (2004).
16. I. Dutta, C. Park and S. Choi, *Mater. Sci. & Engineer.: A,* 379, 401; (2004).
17. B Arfaei, N. Kim and E.J. Cotts, *J. Electron. Mater.,* 41, 362; (2012).
18. I. Dutta, D.Pan, R.A. Marks and S.G. Jadhav, *Mater. Sci. & Engineer.: A,* 410, 48; (2005).
19. B. Talebanpour, U. Sahaym, I. Dutta and P. Kumar, *ASME 2013 Intl.Tech. Conf. & Ex. Pack. Integ. Electron. & Photon. Micros*; (2013).
20. P.T. Vianco, J.A. Rejent and A.C. Kilgo, *J. Electron. Mater.,* 33, 1473; (2004).
21. P.T. Vianco, J.A. Rejent and A.C. Kilgo, *J. Electron. Mater.,* 33, 1389; (2004).
22. I. Dutta, P. Kumar and G. Subbarayan, *JOM,* 61, 29; (2009).
23. M. Hasnine, M. Mustafa, J.C. Suhling, B.C. Prorok, M.J. Bozack and P. Lall, *ECTC, IEEE conference, 2013 IEEE 63rd*; (2013).
24. M.M. Basit, M. Motalab, J.C. Suhling and P. Lall, *ITherm, IEEE Intersociety Conference*; (2014).

25. B. Talebanpour, U. Sahaym and I. Dutta, *IEEE Trans. Device Mater. Reliab.,* 16, 318; (2016).

26. B. Zhou., *MSU Mater. Sci. & Engineer.,* PhD DISSERTATION, 1; (2012).

27. Q. Zhou, B. Zhou, T.R. Bieler, T.K. Lee , *J. Electron. Mater.,* 44, 895; (2015).

28. T.R. Bieler, B. Zhou, L. Blair, A. Zamiri, P. Darbandi, F. Pourboghrat, T.K. Lee and K.C. Liu, *J. Electron. Mater.,* 41, 283; (2012).

29. J. Han, S. Tan, and F. Guo, *J. Electron. Mater.,* 45, 6086; (2016).

30. H. Xu, T.T. Mattila, O. Ratia, M. Paulasto-Kröckel, *ECTC, IEEE conference,* 581; (2010).

31. J.K. Kivilahti and T.T. Mattila, *Trans. Compon. & Pack. Tech.,* 33, (2010).

32. T. Gu, Y. Xu, C.M. Gourlay and T.B. Britton, *Scr. Mater.,* 175, 55; (2020).

33. G. Muralidharan B. Zhou, K. Kurumadalli, C.M. Parish, S. Leslie, Scott, T.R. Bieler, *J. Electron. Mater.,* 43, 57; (2013).

34. H. Jiang, T.R. Bieler, L.P. Lehman, T. Kirkpatrick, E.J. Cotts and B. Nandagopal, *Trans. Compon.& Pack. Tech.* 31, No. 2, 370; (2008).

35. S. Mahin-Shirazi, B.Arfaei, S.Joshi, M.Anselm, P.Borgesen, E.Cotts, J.Wilcox and R. Coyle, *ECTC, IEEE conference*; 976; (2013).

36. F. Mutuku B. Arfaei, R. Coyle, E. Cotts, J. Wilcox, *ECTC, IEEE conference,* 118; (2015).

37. J. Wu, M.S. Alam, K.M.R. Hassan, J.C. Suhling and P. Lall, *ITherm, 19th IEEE Intersociety Conference*; (2020).

38. J.A. Depiver, S. Mallik and E.H. Amalu, *J. Electron. Mater.,* 1; (2020).

39. X. Long, Z. Chen and H. Shi, *21st ICEPT conference*; (2020).

40. M. Abueed, R. Athamenh, J. Suhling and P. Lall, *ITherm, 19th IEEE Conference*; (2020).

41. H. Ma and J.C. Suhling, *J. Mater. Sci.,* 44, 1141; (2009).

42. T. Gu, C.M. Gourlay and T.B. Britton, *J. Electron. Mater.,* 148, 107, (2019).

43. N. Hou, S.A. Belyakov, L. Pay, A. Sugiyama, H. Yasuda and C.M. Gourlay, *Acta Mater.,* 149, 119; (2018).

44. T. Gu, V.S. Tong, C.M. Gourlay and T.B. Britton, *Acta Mater.,* (2020).

45. H.T. Lee and Y.F. Chen, *J. Alloy Compd,* 509, 2510; (2011).

46. H. Esaka, K. Shinozuka and M. Tamura, *Mater. Trans.,* 46, 916; (2005).

47. J.F. Bromley, F. Vnuk and R.W. Smith, *J. Mater. Sci.,* 18, 3143; (1983).

48. M. Kerr and N. Chawla, *Acta Mater.,* 52, 4527; (2004).

49. Z. Mei, D. Grivas, M.C. Shine and J.W. Morris, *J. Electron. Mater.,* 19, 1273; (1990).

50. A. Zamiri, T.R. Bieler and F. Pourboghrat, *J. Electron. Mater.,* 38, 231; (2009).

51. J J. Jiang, T. Zhang, F.P.E. Dunne and T.B. Britton, *Proc. R. Soc. A,* 472, 20150690; (2016).

52. R.D. Doherty, D.A. Hughes, F.J. Humphreys, J.J. Jonas, D.J. Jensen, M.E. Kassner, W.E. King, T.R. McNelley, H.J. McQueen and A.D. Rollett, *Mater. Sci. & Engineer.: A,* 238, 219; (1997).

**Figure Captions**

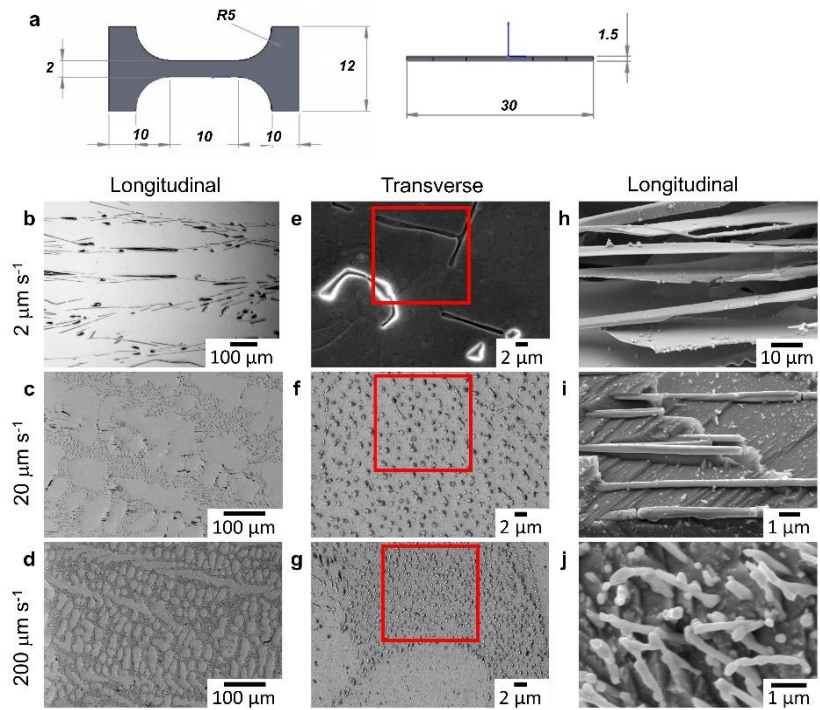

*Figure 1. (a) Schematic diagram of the SAC305 dog-bone samples for creep testing with dimension labelled. (b-j) SEM images of the DS samples at a growth velocity of (b, e, h) 2 μm/s, (c, f, i) 20 μm/s, (d, g, j) 200 μm/s. (b - g) BSE images, (b - d) show the structure of β-Sn dendrites in longitudinal direction (this is the loading direction), where the secondary dendrite arm spacing ($\lambda_2$) is measured (the detail of calculation is given in Supplementary S1), (e - g) show the structure of eutectic region in transverse direction (this is the normal direction) with higher magnification, where the eutectic spacing ($\lambda_e$) is measured within the highlighted red squares from four similar micrographs (the detail of calculation is given in Supplementary S1). (h - j) SE images show the different morphology of IMCs after β-Sn etching.*

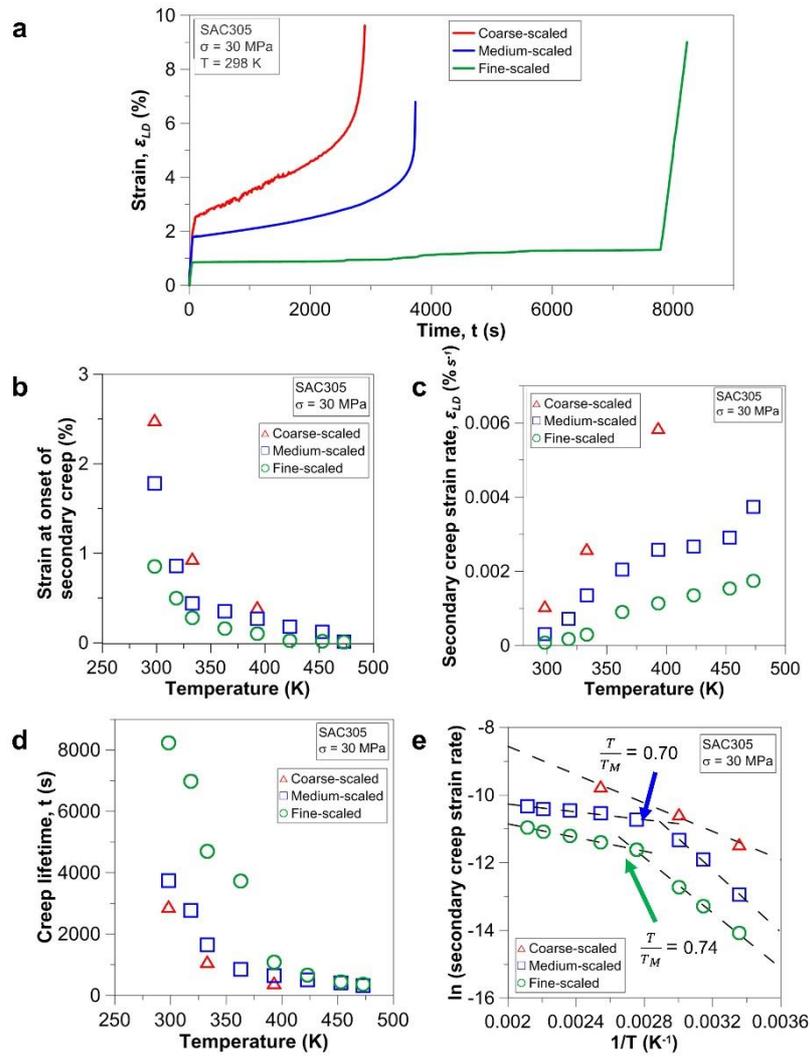

*Figure 2. Comparison of the fixed load (30 MPa) mechanical creep data for the samples with three initial microstructures from 298 to 473 K. The plots show (a) creep curves at 298 K, (b) onset of secondary creep strain vs. temperature, (c) secondary creep strain rate vs. temperature, (d) creep lifetime, (e) creep model analysis of ln (secondary creep strain rate) vs. reciprocal temperature with two creep ranges (high and low temperatures). (The high temperature creep data of each microstructure is included in supplementary Figure S2-3 and in ref [42]).*

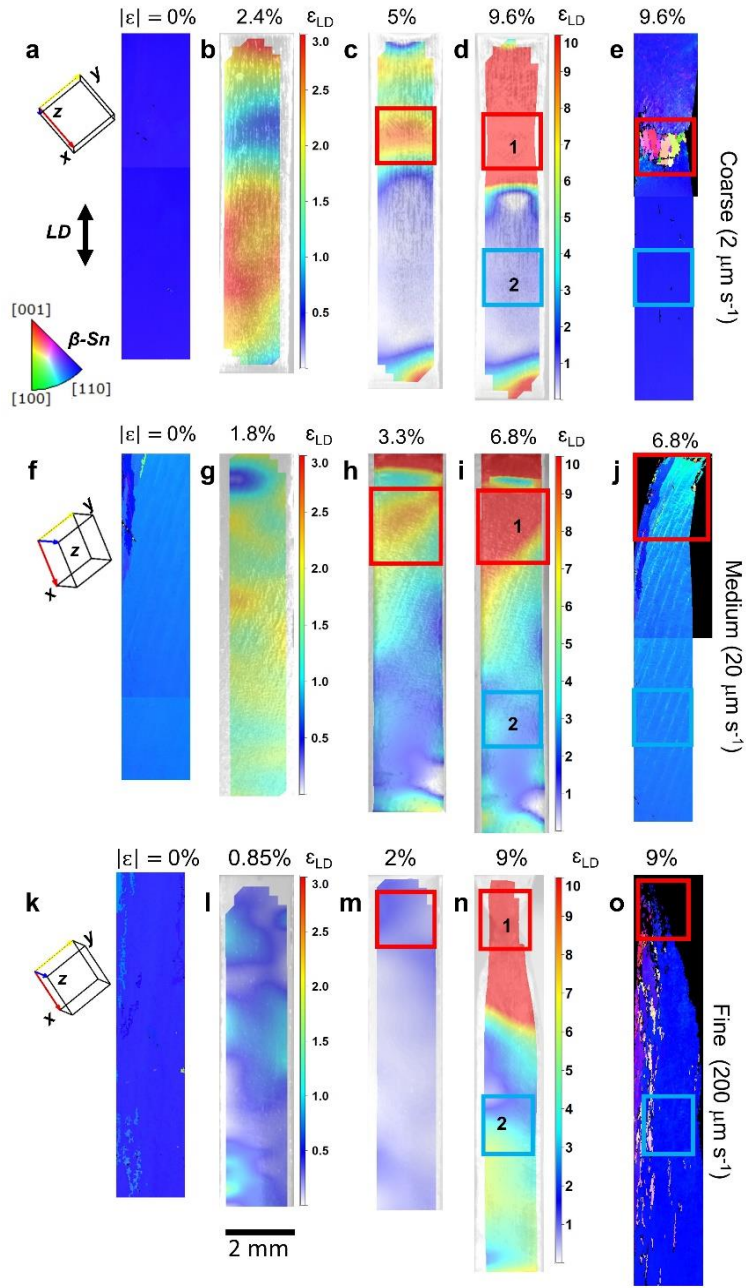

*Figure 3. Comparison for the macroscopic evolution of strain field and microstructure between the three samples with different initial lengthscales. (a-e) Coarse-, (f-j) medium- and (k-o) fine-scaled sample. (b-d, g-i, l-n) are DIC figures and (a, e, f, j, k, o) are EBSD IPF-LD maps. The three samples crept at 298 K with a single near-[110] crystal orientation. (a, f, k) IPF-LD maps at $\varepsilon_{LD} = 0\%$ with main crystal orientation shown. The DIC figures (b, g, l) at onset of secondary creep stage ($\varepsilon_{LD} = 2.4\%$, 1.8% and 0.85% respectively), (c, h, m) near onset of tertiary creep stage (fracture) ($\varepsilon_{LD} = 5\%$, 3.3% and 2% respectively), (d, i, n) at the end of tertiary creep stage ($\varepsilon_{LD} = 9.6\%$, 6.8% and 9% respectively). (g) IPF-LD map at $\varepsilon_{LD} = 9.6\%$, 6.8% and 9% (fracture).*

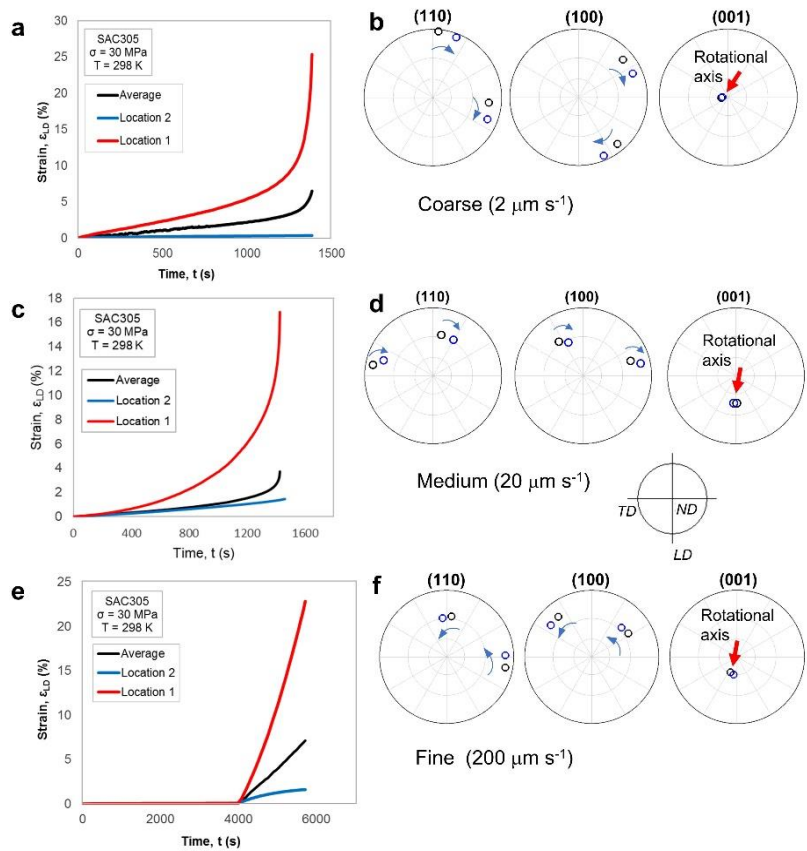

*Figure 4. (a, c, e) Creep curves from secondary to tertiary creep stage showing the heterogeneity in strain development through the lengthscale (mm) of the sample. Location 1 (red line) and location 2 (blue line) are extracted from Figure 3 (d, i n) in the corresponding locations, which are compared to the average creep curves of the samples (black line). Change in orientations of the samples is given in the PF (b, d, f), where the initial orientations are black circles and the rotated orientations (after fracture) are blue circles.*

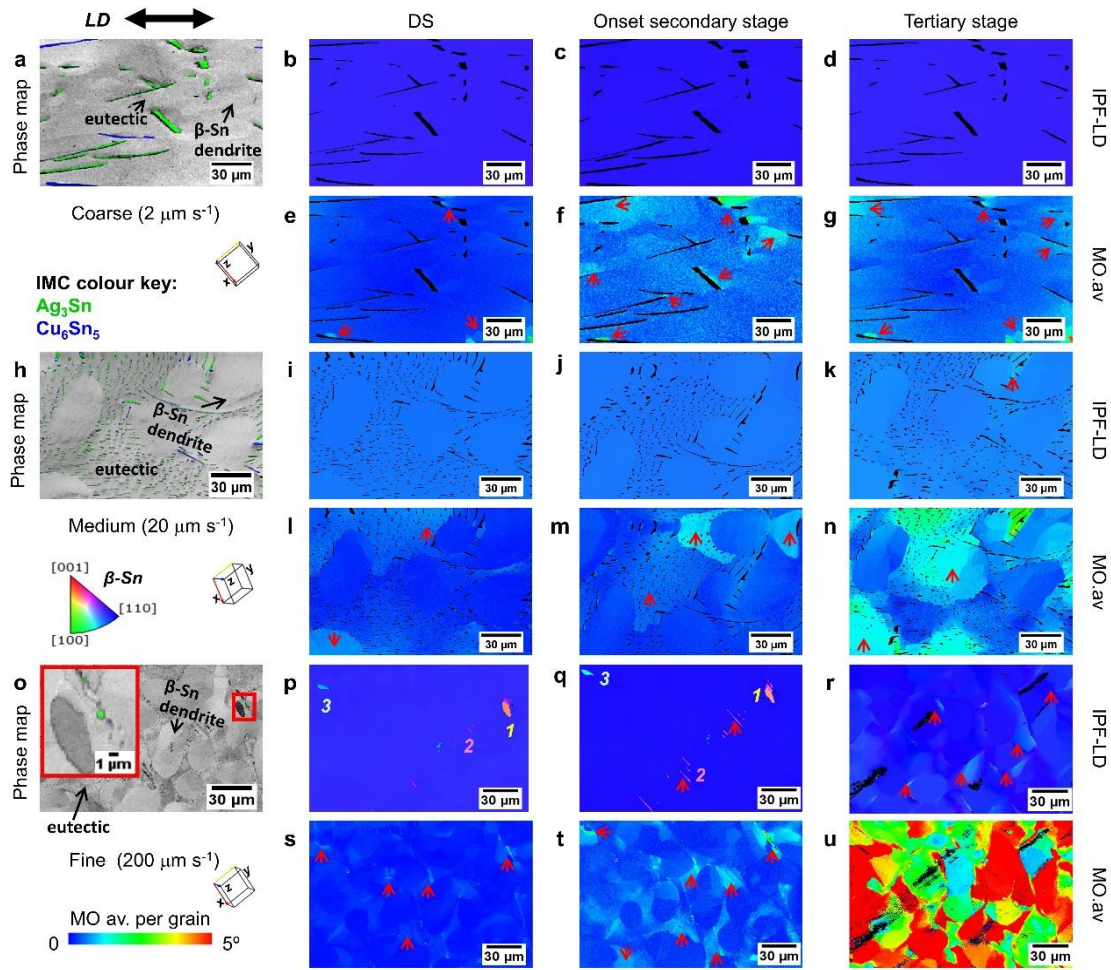

*Figure 5. Higher magnification EBSD crystal orientation maps showing microstructural evolution for the three samples within the uniformly-deformed region (location 2 in Figure 3b, 3g, 3i). (a - g) Coarse-, (h - n) medium-, (o - u) fine-scaled sample. (a, b, e, h, i, l, o, p, s) at DS condition, (c, f, j, m, q, t) at onset of secondary creep, (d, g, k, n r, u) at the end of tertiary creep. (a, h, o) Phase map overlaid on pattern quality map showing the dendritic and eutectic regions with IMCs highlighted in green ($Ag_3Sn$) and blue ($Cu_6Sn_5$). (b - d, i - k, p - r) IPF-LD maps. (e - g, l - n, s - u) Misorientation average (MO av.) maps. Here only the β-Sn phase has been indexed and a common misorientation angle with respect to 'grain' as identified with EBSD thresholding of grain boundaries at 5°.*

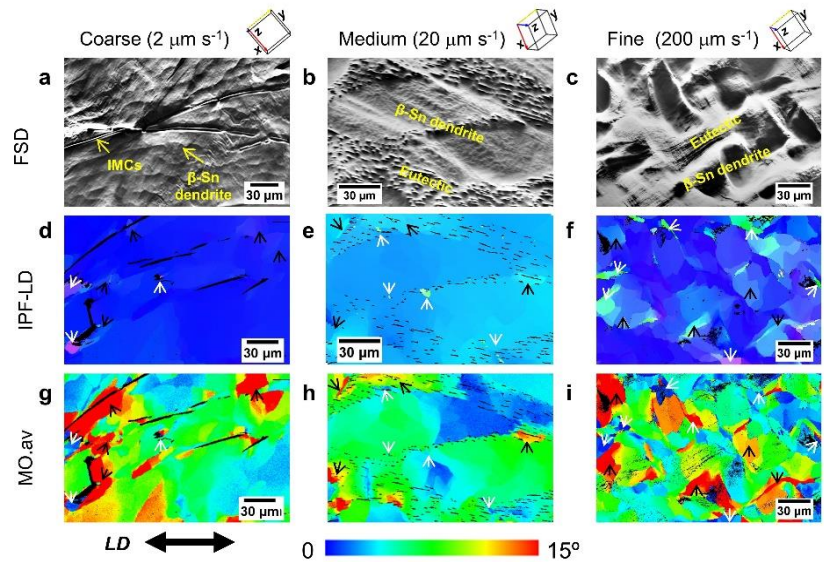

*Figure 6. Higher magnification EBSD crystal orientation maps showing microstructural evolution for the three samples within the necked regions (location 1 in Figure 3e, 3j, 3o) at the end of tertiary creep. (a - c) FSD images showing the morphology of the samples with dendrite and eutectic regions labelled. (d - f) IPF-LD maps where the formed recrystallised grains are indicated with white arrows. (g - i) MO av. maps where the MO hot spots are indicated with white arrows. Here only the β-Sn phase has been indexed.*

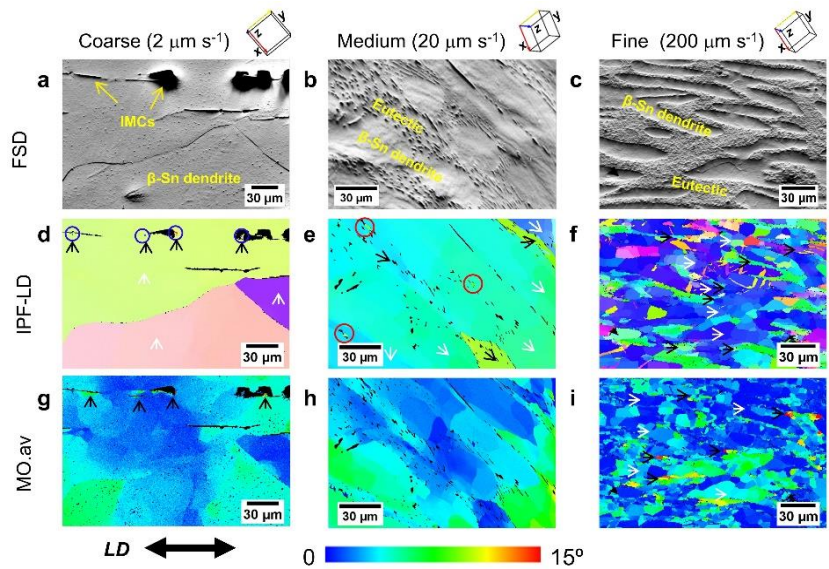

*Figure 7. Higher magnification EBSD crystal orientation maps showing microstructural evolution for the three samples at the fracture surfaces (location 1 in Figure 3e, 3j, 3o) at the end of tertiary creep (the maps are taken from the polished surface after deformation). (a - c) FSD images showing the morphology of the samples with dendrite and eutectic regions labelled. (d - f) IPF-LD maps with significant recrystallisation. (g - i) MO av. maps show significant increase in misorientation. Here only the β-Sn phase has been indexed.*

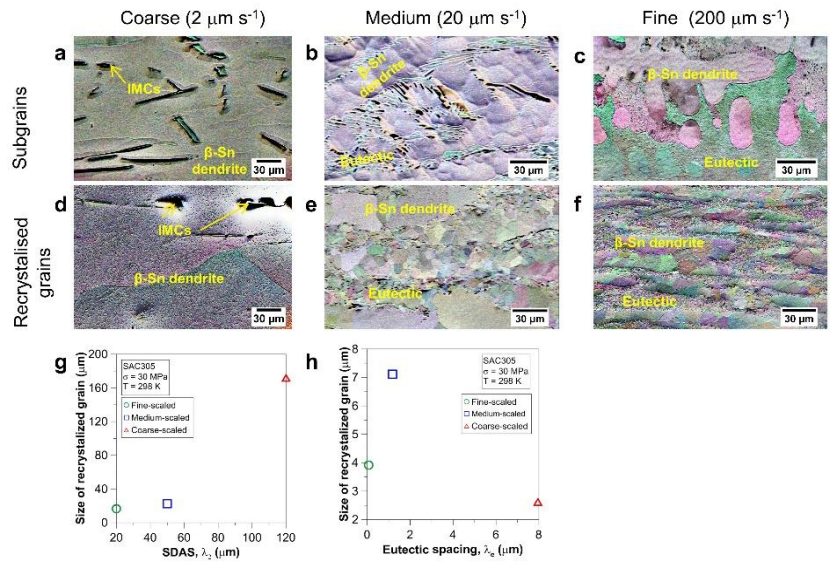

*Figure 8. Comparison of the size for subgrain and recrystallised grains between the three samples with different initial microstructures. (a-f) FSD images showing the morphology of the β-Sn dendrite and IMC together with the orientation contrast between subgrains and recrystallised grains. The subgrains (a - c) and recrystallised grains (d - f) are illustrated at tertiary creep within the uniformly-deformed and highly-strained regions of the sample respectively. (g, h) The plots give the size of the recrystallised grains vs. $\lambda_2$ (g) and $\lambda_e$ (h).*

30